\documentclass[slac_one]{revtex4}
\usepackage{graphicx}
\usepackage{fancyhdr}
\usepackage{color}
\def\reffi#1{\mbox{Fig.~\ref{#1}}}

\def\refta#1{\mbox{Tab.~\ref{#1}}}
\def\citere#1{\mbox{Ref.~\cite{#1}}}

\def\mathswitch#1{\relax\ifmmode#1\else$#1$\fi}
\def\mathswitchr#1{\relax\ifmmode{\mathrm{#1}}\else$\mathrm{#1}$\fi}
\newcommand{\Ph}{\mathswitchr h}

\newcommand{\PH}{\mathswitchr H}

\newcommand{\mh}{\mathswitch {m_\Ph}}

\newcommand{\MH}{\mathswitch {M_\PH}}
\newcommand{\lsim}
{\;\raisebox{-.3em}{$\stackrel{\displaystyle <}{\sim}$}\;}

\newcommand{\gev}{\,\, \mathrm{GeV}}

\newcommand{\cp}{{\cal CP}}

\begin{document}

\definecolor{Blue}{named}{Blue}
\definecolor{Red}{named}{Red}

\title{{\small{2005 International Linear Collider Workshop - Stanford,
U.S.A.}}\\ 
\vspace{12pt}
The LHC and the ILC} 

%

\author{G.~Weiglein}
\affiliation{IPPP, University of Durham, Durham DH1 3LE, U.K.}

\begin{abstract}
The synergy between the Large Hadron Collider and the International
Linear Collider during concurrent running of the two machines has the 
potential to maximise the physics gain from both facilities. Some
examples of detailed case studies of the interplay between the LHC and
ILC are given, with a particular emphasis on new results that have been
obtained after the first LHC / ILC Study Group report was released.
\end{abstract}

\maketitle

\thispagestyle{fancy}


\section{INTRODUCTION}

Ground-breaking discoveries are expected from the experiments under 
construction at the Large Hadron Collider (LHC) and those planned for 
the International Linear Collider (ILC).
These high-energy particle accelerators will open up a new energy
domain that will allow us to examine the very fabric of matter, 
space and time. 

The LHC and ILC will probe this new TeV energy regime in very different 
ways, as a consequence of the distinct features of the
two machines. Due to its
high collision energy and luminosity, the LHC has a large mass reach for
the discovery of new heavy particles. The striking advantages of the ILC
are its clean experimental environment, polarized beams, and tunable
collision energy. The ILC can thus perform precision measurements and
detailed studies of directly accessible new particles, and also has
exquisite sensitivity to quantum effects of unknown
physics. Indeed, the fingerprints of very high-scale new physics (e.g.\
very high mass particles) will often only be manifest in small effects whose
measurement requires the greatest possible precision.

The LHC, currently under construction at CERN (Geneva), is
scheduled to go into operation in 2007. A timely realisation of the ILC,
leading to a start
of data taking by the middle of the next decade, would give rise to a
significant period of overlapping running with the LHC.
The complementarity between hadron and lepton colliders has often led
to a concurrent operation of these two types of machines in the past,
and there is a long history of productive synergy between them.
The exploration of the TeV energy regime should 
give us clues on how particles obtain the property of mass, whether the
different forces that we experience
in nature are in fact different manifestations of only one fundamental
force, whether space and time are embedded into a wider framework of
supersymmetric coordinates, whether dark matter can be produced in the
laboratory, and on many other fundamental questions. Therefore an
even greater synergy can be expected from the LHC and
ILC compared to previous generations of hadron and lepton colliders.
While each machine has its own independent and complementary physics
agenda, there are expected to be significant scientific advantages in
having them operate at the same time: with combined analyses of the data
during concurrent running of both machines, the results obtained at one
machine can directly influence the way analyses are carried out at the
other machine, leading to optimised experimental strategies and
dedicated searches. 

In order to assess the prospective synergy of concurrent running of the
LHC and the ILC in a quantitative way, detailed informations 
on the experimental capabilities of the LHC and ILC in different
scenarios of physics within and beyond the Standard Model (SM) are
needed as input. Studying the interplay between the LHC and ILC requires 
close collaboration of experts from the LHC and ILC as
well as from theorists and experimentalists. A world-wide working group,
the LHC / ILC Study Group, has formed as a
collaborative effort of the hadron collider and linear collider
experimental communities and theorists 
with the aim to explore the interplay between
the LHC and ILC. A first working group report has recently been
completed~\cite{lhclc}. Many different scenarios of physics at the
TeV scale have been investigated in this report. Starting from an
assessment of the prospective experimental input from the LHC and ILC in
each scenario, the possible synergy from concurrent operation of LHC and
ILC has been studied. For scenarios where detailed experimental
simulations of the possible measurements and the achievable accuracies
are available both for the LHC and ILC, the LHC / ILC interplay has been
investigated in a quantitative manner. In other scenarios the
prospective physics gain arising from the LHC / ILC interplay has 
been discussed in a qualitative way. It has been demonstrated that 
the synergy between the LHC and ILC will extend the physics potential 
of both machines. Results from both
colliders will be crucial in order to identify the
underlying physics in the new territory opening up at the TeV scale.

The first working group report~\cite{lhclc} has shown that the 
interplay between the LHC and ILC is a very rich field, of which only very
little has been explored so far. Dedicated studies, 
taking into account information from both the LHC and the
ILC in a coherent framework, are necessary in order to determine the
expected physics gain. The detailed experimental simulations
currently carried out by the ATLAS and CMS collaborations in preparation 
for the start of data taking at the LHC together with the ongoing
efforts of the
ILC physics groups provide an ideal input for future studies of LHC / ILC
interplay.

The possible physics gain from concurrent running of the LHC and ILC
has found a lot of interest in the wider scientific community and among 
funding agencies~\cite{itrp,ippppparc,epp}.
In the U.S., for instance, the High Energy Physics
Advisory Panel (HEPAP) of the U.S.\ Department of Energy has formed a
subpanel that has the charge to address the complementarity between the
LHC and the ILC~\cite{hepap}. The EPP Decadal Survey panel, with the 
charge to identify, articulate and prioritise the scientific questions and
opportunities that define elementary particle physics, has posed several
questions to the scientific community that are specifically related to
concurrent running of the LHC and ILC~\cite{eppbagger}. 
The above examples indicate that there exists a high demand 
for a quantitative account of the expected physics gain of concurrent
operation of both facilities. A continuing effort is therefore required
in order to provide the results necessary for a well-founded judgement
of the interplay between the LHC and ILC.

In the following, some recent results on LHC / ILC interplay are
summarised. Particular emphasis is put on new results that have been
obtained after the first LHC / ILC Study Group report has been
completed.

\section{SOME RECENT RESULTS ON LHC / ILC INTERPLAY}

\subsection{Determination of Higgs-boson couplings}

If one or more new particles are observed in a way that is consistent
with Higgs-boson production, it will be of utmost importance to
precisely determine as many properties of the new particle(s) as
possible. Ratios of couplings of a light SM-like Higgs boson can be
measured at the LHC in a fairly model-independent
way~\cite{ATL-PHYS-2003-030}, while mild theory assumptions are
necessary in order to extract absolute values for the
couplings~\cite{higgscoupllhc,higgscouplprev}. The use of theory
assumptions can be avoided by using information from the ILC in the LHC
analysis.
This has been studied for the example
of the Yukawa coupling of the Higgs boson to a pair of top quarks.
In the first phase of the ILC with a centre of mass energy of about
500~GeV the $t \bar th$ coupling can only be measured with limited precision
for a light Higgs boson $h$, as a consequence of the phase space
suppression of the $e^+e^- \to t \bar th$ production process. The LHC will 
provide a measurement
of the $t \bar th$ production cross section times the decay branching
ratio (for $h \to b \bar b$ or $h \to W^+W^-$). The ILC, on the other hand,
will perform precision measurements of the decay branching ratios.
Combining LHC and ILC information will thus allow one to extract the top
Yukawa coupling. This has been demonstrated in \citere{topyukds}, where 
an accuracy on the $g_{\rm t \bar t h}$ coupling of 15--20\% has been found
for $\mh \lsim 200$~GeV. 

\begin{figure*}[th]
\centering
\includegraphics[width=0.45\textwidth]{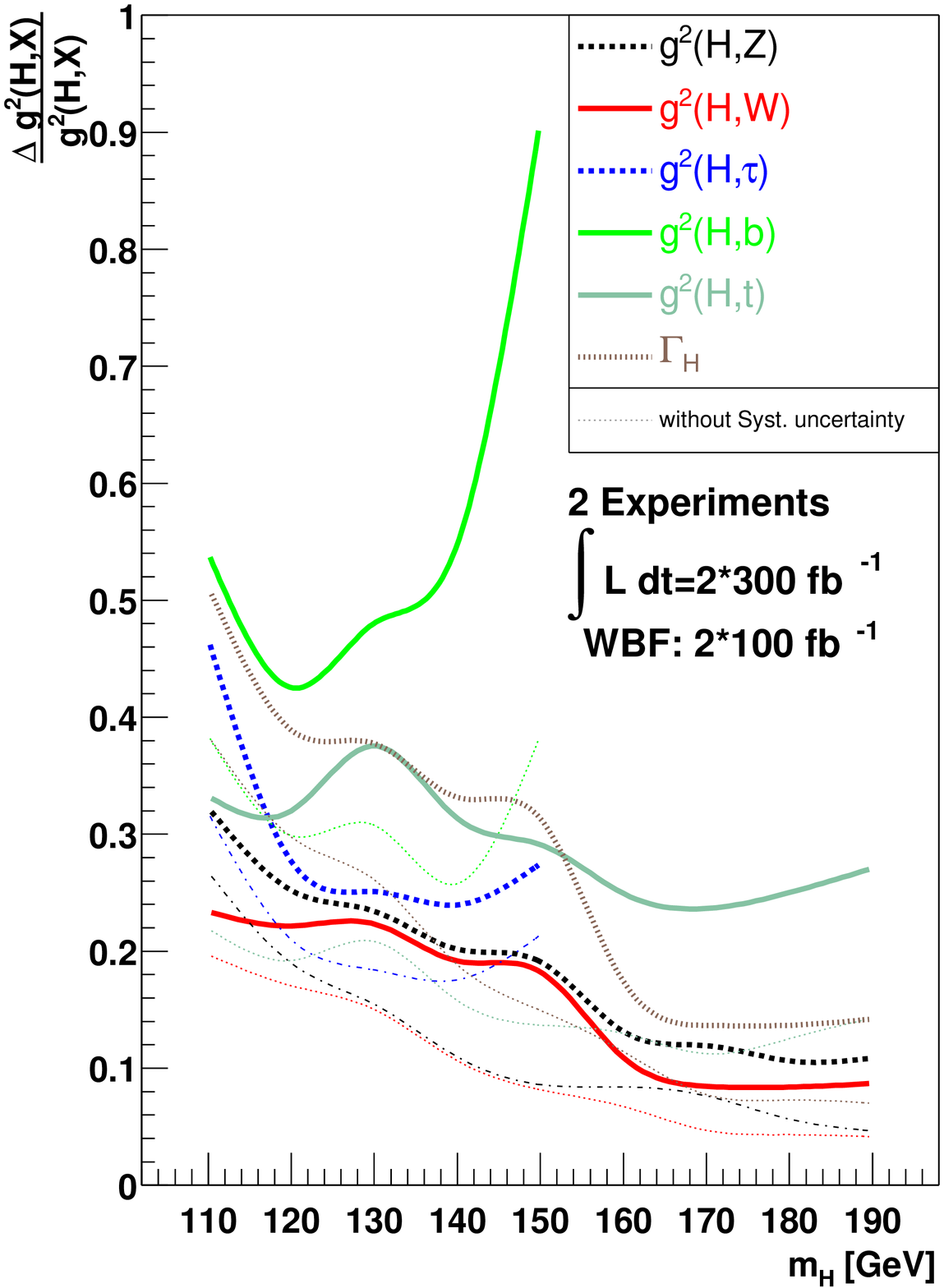}
\includegraphics[width=0.45\textwidth]{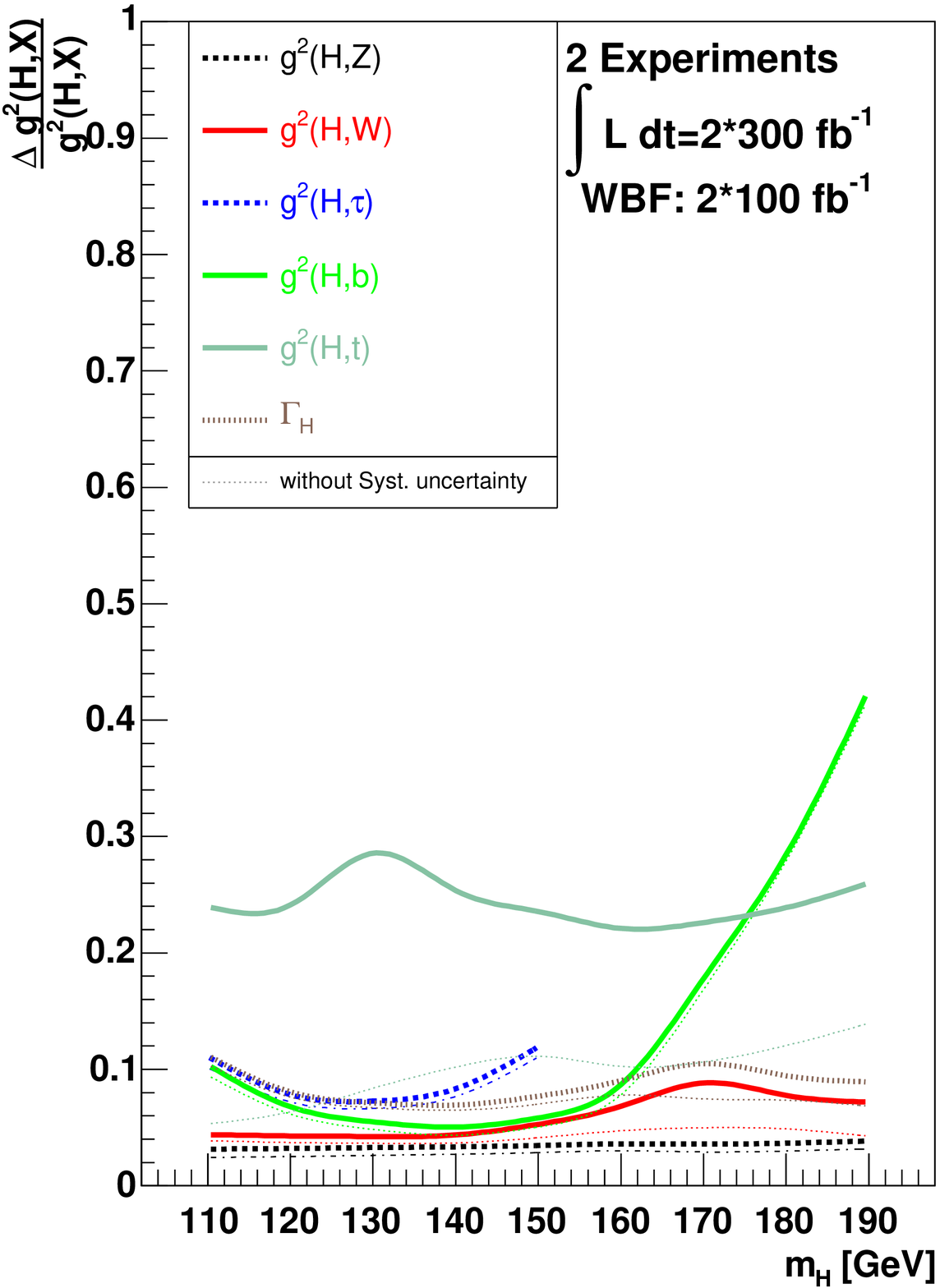}
\caption{Relative accuracies of squared Higgs-boson couplings and the
total Higgs-boson width achievable at the LHC alone with mild theory
assumptions~\cite{higgscoupllhc} (left plot) and in a combined analysis
using input from the LHC and ILC~\cite{higgscoupllhcilc} (right plot).} 
\label{fig:higgscoupl}
\end{figure*}

The analysis of \citere{topyukds} has been
extended in \citere{higgscoupllhcilc}, where a combined fit has been
performed that includes all LHC channels and ILC input on the
Higgs-boson mass, branching ratios and production cross sections.
The results are shown in \reffi{fig:higgscoupl}, where the result for
the LHC with mild theory assumptions (left plot) is compared with the
result of the combined LHC / ILC model-independent analysis (right
plot). The ILC input leads to a drastic improvement in the accuracy of the 
coupling determination. For most couplings the combined analysis is 
completely driven by the precision achievable at the ILC alone.
For the $t \bar th$ coupling (and also for the $h \gamma\gamma$
coupling, which is not shown in \reffi{fig:higgscoupl}), on the other
hand, the combined analysis improves over the analyses both at the LHC
alone and at the ILC alone. The resulting accuracy on 
the $g_{\rm t \bar t h}$ coupling in \reffi{fig:higgscoupl} is 11--14\%
(note that \reffi{fig:higgscoupl} shows the squared coupling 
$\Delta g_{\rm t \bar t h}^2/g_{\rm t \bar t h}^2$ rather than
the coupling itself), 
which corresponds to a relative improvement of about 30\% compared to the
analysis in \citere{topyukds}.

The complementarity of the LHC and the ILC in both the $e^+e^-$ and the 
photon collider mode for determining Higgs-boson couplings has been 
investigated in \citere{higgsphot}. The decays $H \to WW, ZZ$ have been
analysed for $200 \gev \lsim \MH \lsim 350 \gev$ in a Two Higgs Doublet
Model (II) with $\cp$ violation. It has been found that the
measurements at the photon collider are complementary to the ones at
the LHC and the ILC, since they are sensitive to different combinations
of Higgs-boson couplings. A combined analysis of the LHC and the ILC in
its $e^+e^-$ and photon collider modes is necessary in order to precisely
determine the $\cp$-violating $H$--$A$ mixing angle $\phi_{\rm HA}$.

\subsection{Higgs physics in the NMSSM}

While the search for the Higgs boson of the SM has been well studied at
the LHC and ILC, physics beyond the SM can give rise to very different 
Higgs phenomenology. This can be due to modified couplings, mixing with
other states, non-standard production processes or Higgs decays into new
particles. Even in the Minimal Supersymmetric Extension of the Standard
Model (MSSM) the Higgs-boson properties can be rather different from
those of a SM Higgs boson. The Next-to-Minimal Supersymmetric Model
(NMSSM) has recently found considerable attention as an attractive
extension of the MSSM, since it allows to avoid the fine-tuning
and ``little hierarchy'' problems of the $\cp$-conserving
MSSM. While the NMSSM is theoretically well motivated,
its Higgs phenomenology at the LHC can be very challenging~\cite{nmssmhiggs}. 
This is due to the fact that over a large part of the parameter space
the SM-like $\cp$-even Higgs boson of the NMSSM dominantly decays into
two light $\cp$-odd Higgs bosons, $h\to aa$. Confirmation of the nature
of a possible LHC signal at the ILC would be vital. For example,
the $WW\to h\to aa$ signal, as well
as the usual $e^+e^-\to Z h\to Z aa$ signal,
will be highly visible at the
ILC due to its cleaner environment and high luminosity. The ILC will
furthermore be able to measure important properties of the $\cp$-odd
scalar. Even if a trustworthy signal is seen at the LHC, the ILC will
probably be essential to determine that the signal observed at the LHC
indeed corresponds to a Higgs boson.

If no clear Higgs signal has been established at the LHC, it
will be crucial to investigate with the possibilities of the ILC whether
the Higgs boson has not been missed at the LHC because of its
non-standard
properties. This will be even more the case if the gauge sector does not
show indications of strong electroweak symmetry breaking dynamics. The
information obtained from the ILC can therefore be crucial for
understanding the physics of mass generation.
The particular power of the ILC is its ability to look for 
$e^+e^- \to ZH$
in the inclusive $e^+e^- \to ZX$ missing-mass distribution
recoiling against the $Z$ boson. Even if the Higgs boson decays
completely
invisibly or different Higgs signals overlap in a complicated way, the
recoil mass distribution  will reveal the Higgs boson mass spectrum
of the model.

Another challenging NMSSM scenario is a singlet dominated light Higgs.
While this state has reasonably large production cross sections at the
LHC, it would be difficult to detect as it mainly decays hadronically.
Such a state could be discovered at the ILC. From the measurement of its
properties, the masses of the heavier Higgs bosons could be predicted,
guiding in this way the searches at the LHC. For a very heavy singlet
dominated Higgs state, on the other hand, the kinematic reach of the
LHC will be crucial in order to verify that a non-minimal Higgs sector is
realised. Thus, input from both the LHC and the ILC will be needed in
order to provide complete coverage over the NMSSM  parameter space.

\subsection{Supersymmetry at the LHC and ILC}

The production of
supersymmetric particles at the LHC will be dominated by the production
of coloured particles, i.e.\ gluinos and squarks.
Searches for the signature of jets and missing energy at the LHC will cover
gluino and squark masses of up to 2--3~TeV. The main
handle to detect uncoloured particles will be
from cascade decays of heavy gluinos and squarks, since in most 
scenarios of supersymmetry (SUSY) the uncoloured particles are lighter
than the coloured ones. An example of a possible decay chain is
$\tilde g \to \bar q \tilde q \to \bar q q \tilde\chi_2^0 \to
\bar q q \tilde\tau \tau \to \bar q q \tau\tau \tilde\chi_1^0$,
where $\tilde\chi_1^0$ is assumed to be the lightest supersymmetric
particle (LSP).
Thus, fairly long decay chains giving rise to the production of several
supersymmetric particles in the same event and leading to rather
complicated final states can be expected to be a typical feature of SUSY 
production at the LHC. In fact, the main background for measuring SUSY
processes at the LHC will be SUSY itself.

The ILC, on the other hand, has good prospects for the production of
the light 
uncoloured particles. The clean signatures and small backgrounds at the
ILC as well as the possibility to adjust the energy of the collider to
the thresholds at which SUSY particles are produced will allow a precise
determination of the mass and spin of supersymmetric particles and of
mixing angles and complex phases.

In order to establish SUSY experimentally, it will be necessary to
demonstrate that every particle has a superpartner, that their spins
differ by $1/2$, that their gauge quantum numbers are the same, that
their couplings are identical and that certain mass relations hold.
This will require a large amount of experimental information, in
particular precise measurements of masses, branching ratios,
cross sections, angular distributions, etc. A precise knowledge of as
many SUSY parameters as possible will be necessary to disentangle the
underlying pattern of SUSY breaking and to reveal a possible SUSY nature
of dark matter~\cite{darkmatter}. 

It has been demonstrated that the analysis of SUSY particle production
at the LHC can benefit very significantly from experimental results
obtained at the ILC~\cite{lhclc}.
As mentioned above, at the LHC the dominant production mechanism is pair 
production of
gluinos or squarks and associated production of a gluino and a squark. 
For these processes, SUSY particle masses normally 
have to be determined from the 
reconstruction of long decay chains which end in the production of the
LSP. The invariant
mass distributions of the observed decay products exhibit thresholds and 
end-point structures. The kinematic structures can in turn be expressed
as a function of the masses of the involved supersymmetric 
particles. The LHC is
sensitive in this way mainly to mass {\em differences}, resulting in a
strong correlation between the extracted particle masses. In particular,
the LSP mass is only weakly constrained. This uncertainty propagates
into the experimental errors of the heavier SUSY particle masses.
Furthermore, the determination of the masses from the end-point
expressions in general leads to ambiguities~\cite{gjelsten}. A specific
set of end-point values can often be produced by several sets of masses,
leading to multiple solutions in a $\chi^2$ fit for determining the
masses even in the favourable case of the SPS~1a benchmark
scenario~\cite{sps}. 
The precision measurements at the ILC of the colour--neutral part of the SUSY
particle spectrum, in particular of the LSP mass, eliminate
a large source of uncertainty in the LHC analyses. The ambiguities in
the LHC analyses can be resolved using ILC input on the LSP mass and the
slepton masses~\cite{gjelsten}. Inserting the precision measurement of
the LSP mass at the ILC into the LHC analyses furthermore leads
to a substantial improvement in
the accuracy of the reconstructed masses of the particles in the decay
chain~\cite{lhclc,gjelsten}.

In general ILC input will help to significantly reduce the model
dependence of the LHC analyses. Intermediate states that appear in the
decay chains detected at the LHC can be produced directly and
individually at the ILC. Since their spin and other properties can be
precisely determined at the ILC, it will be possible to unambiguously
identify the nature of these states as part of the SUSY spectrum. In
this way it will be possible to verify the kind of decay chain observed at the
LHC. 
Once the particles in the lower parts of the decay cascades have
been clearly identified, one can include the MSSM predictions for their
branching ratios into a constrained fit. This can be helpful in order
to determine the couplings of particles higher up in the decay chain.

\begin{figure*}[thb]
\centering
\resizebox{.7\textwidth}{!}{
\setlength{\unitlength}{1em}
\begin{picture}(14,14)
\includegraphics[width=14.0em]{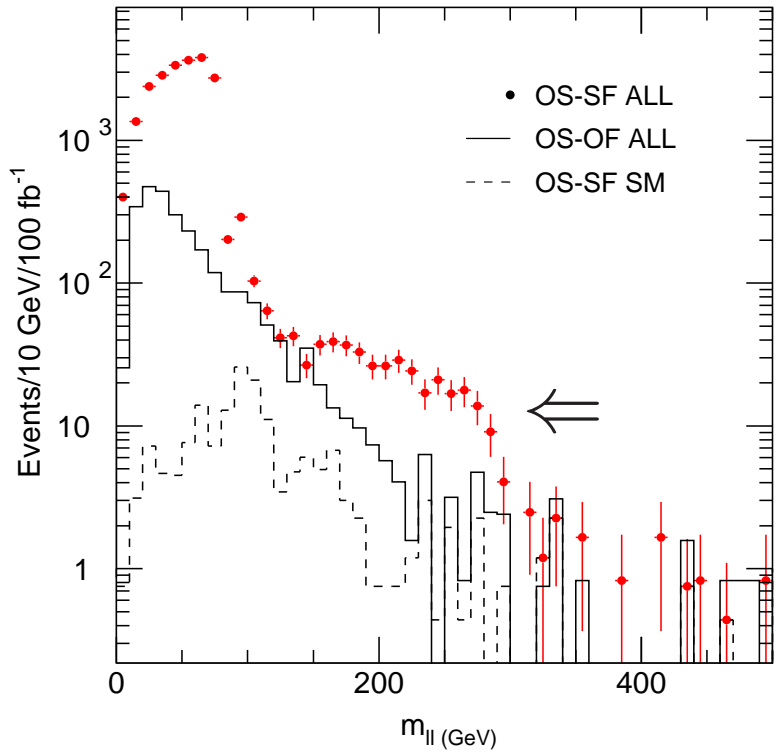}
\put(-5.4,5.8){{\large $\Leftarrow$}}
\end{picture}
}
\caption{Simulation of invariant mass spectra at the LHC 
in the SPS~1a scenario~\cite{dkmnpJHEP}:
opposite-sign same-flavour (OS-SF) leptons total (full dots), opposite-sign
opposite-flavour (OS-OF) leptons total (solid line) and opposite-sign
same-flavour leptons in the SM (dashed line). The signals of
$\tilde\chi^0_2$ and $\tilde\chi^0_4$ consist of OS-SF leptons. The
arrow indicates the edge associated with the decay of $\tilde\chi^0_4$
that can be identified using the ILC prediction 
$m_{\tilde\chi^0_4} = 378.3 \pm 8.8$~GeV.} 
\label{fig:dkmnp}
\end{figure*}

A detailed study of the interplay of the LHC and ILC in the gaugino
sector has been performed in \citere{dkmnpJHEP}. In this analysis,
carried out in the SPS~1a scenario~\cite{sps}, the 
measurements of the masses of the two lightest neutralinos, the lighter
chargino, the selectrons and the sneutrino at the ILC are used to
determine all parameters in the neutralino and chargino sector. This
allows then to predict
the properties of the heavier neutralinos. For the heaviest neutralino, 
$\tilde\chi^0_4$, the ILC measurements give rise to the prediction
$m_{\tilde\chi^0_4} = 378.3 \pm 8.8$~GeV for its mass. 
It was demonstrated in \citere{dkmnpJHEP} that the ILC 
input makes it possible to identify the heaviest neutralino at the LHC,
see \reffi{fig:dkmnp},
and to measure its mass with high precision. Feeding this information back
into the ILC analysis significantly improves the determination of the 
fundamental SUSY parameters from the neutralino and chargino sector at 
the ILC. 

The described analysis is a typical example of LHC / ILC synergy.
If a statistically not very pronounced (or even marginal)
signal is detected at the LHC, input from the ILC can be crucial in order
to identify its nature. In fact, the mere existence of an ILC prediction
as input for the LHC searches increases the statistical sensitivity of
the LHC analysis. This happens since a specific hypothesis is tested,
rather than performing a search over a wide parameter space. In the
latter case, a small excess {\em somewhere\/} in the parameter space is
statistically much less significant, since one has to take into account
that a statistical fluctuation is more likely to occur in the
simultaneous test of many mass hypotheses. 
Beyond the enhancement of the statistical sensitivity, predictions based
on ILC input can also give important guidance for dedicated searches at
the LHC. This could lead to an LHC analysis with optimised cuts or even
improved triggers. ILC input might also play an important role in the
decision for upgrades at later stages of LHC running.

\begin{figure*}[htb]

\vspace{.7cm}

\setlength{\unitlength}{1cm}
\mbox{} \hspace{-11cm}
\resizebox{16cm}{!}{
\begin{picture}(10,5)
\put(9.15,3.5){\small{\color{red} $\tilde{\chi}^0_3$\quad}
and {\color{blue}\quad $\tilde{\chi}^0_4$}}
\put(7.5,-.5){\small gaugino character of
{\color{red}$\tilde{\chi}^0_3$},
{\color{blue}$\tilde{\chi}^0_4$}}
\put(5,4.5){\small $m_{\tilde{\chi}^0_i}$/GeV}
\put(4.5,1.2){\small LHC:}
\put(4.5,.9){\small measurement}
\put(4.95,.6){\small of
\color{green}$m_{\tilde{\chi}^0_i}\rightarrow$}
\put(6,-1){\includegraphics[width=70mm]{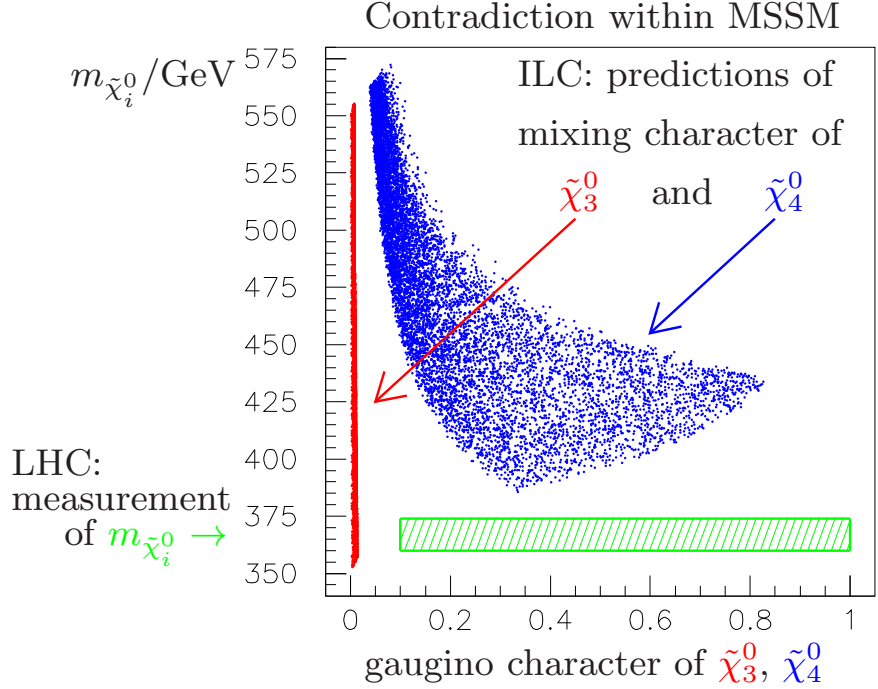}}
\put(7.5,5){\small Contradiction within MSSM}
\put(8.8,4.5){\small ILC: predictions of}
\put(8.8,4){\small  mixing character of}
\end{picture}
}

\vspace{.7cm}

\caption{
Predicted masses and mixing character for the heavier neutralinos
$\tilde{\chi}^0_3$  and $\tilde{\chi}^0_4$ within the MSSM based on the 
measurements of the light neutralino and chargino states at the ILC 
with 500~GeV c.\ m.\ energy. Detection of the neutralinos in cascade
decays at the LHC requires a sufficient 
gaugino admixture (in the plot a lower bound of about 10\% gaugino
admixture has been applied). The indicated LHC measurement (with ILC
input) of $m_{\tilde{\chi}^0}=367\pm 7$~GeV is therefore inconsistent
within the MSSM, but consistent with the NMSSM prediction.
\label{fig:mausnmssm}
}
\end{figure*}

On the other hand, if the observations at the LHC are not consistent
with the predictions from ILC input within the MSSM, 
this would be an important hint that the observed particles cannot be
consistently described within the minimal model. As a recent example, in 
\citere{mausnew} an NMSSM scenario has been studied where the
light neutralinos have a significant singlino component. This scenario 
cannot be distinguished from the MSSM by cross section and mass 
measurements, since the Higgs sector and the light neutralino / chargino
spectra and cross sections are almost identical in the two
models~\cite{mausnew}. The parameter determination from the light
neutralino and chargino states at the ILC with 
500~GeV c.\ m.\ energy (and at the LHC alone) 
could therefore be carried out in this
scenario as in the MSSM and would not lead to any contradictions.
As discussed above, the ILC input will allow to predict the properties
of the heavy neutralinos, $\tilde\chi^0_3$ and $\tilde\chi^0_4$. 
In this scenario, the ILC measurements predict within the
MSSM an almost pure higgsino-like state for $\tilde\chi^0_3$ and 
a mixed gaugino-higgsino-like state for $\tilde\chi^0_4$, see 
\reffi{fig:mausnmssm}. An almost pure higgsino-like $\tilde\chi^0_3$
would not have sufficiently large couplings to squarks and would
therefore not be detectable at the LHC. The detection of $\tilde\chi^0_3$
at the LHC would therefore be inconsistent within the MSSM, but in
agreement with the NMSSM prediction, see \reffi{fig:mausnmssm}. 
The combined LHC / ILC analysis
would therefore allow to distinguish between the MSSM and non-minimal
supersymmetric models.

In order to establish SUSY experimentally and to determine the
SUSY-breaking patterns, it is necessary to accurately determine as many
Lagrangian parameters as possible. Since most observables depend on a variety 
of parameters, one will have to perform a global
fit~\cite{sfitter,fittino} of the SUSY model
to a large number of experimental observables. As the measurements at the 
LHC and the ILC in general probe different sectors of the MSSM Lagrangian, 
the combination of LHC and ILC data will be crucial in order to obtain
comprehensive information on the underlying structure of the model. 
Attempts to fit only individual sectors of the theory turned out to be
unsuccessful~\cite{lhclc,fittino}.
In \citere{fittino} it has
been demonstrated that only the combination of measurements of both the
LHC and the ILC offers a complete picture of the MSSM model parameters in
a reasonably model independent framework (and without employing
theoretical assumptions about a specific SUSY-breaking scenario). In a
combined fit based on input from the LHC and the ILC in a variant of the
SPS~1a benchmark scenario~\cite{sps} 19 Lagrangian
parameters (neglecting complex phases and mixing between the
generations) have been determined. The input comprised mass measurements 
of supersymmetric particles at the LHC and ILC, using as experimental
uncertainties the values obtained in \citere{lhclc}, the cross section
measurement of the Higgs-strahlung production of a light Higgs at the
ILC, cross section times branching fraction measurements at the ILC for
all processes with a sufficiently high rate, Higgs-boson branching
fractions measured at the ILC and ratios of branching fractions
obtained at the LHC (as discussed above, absolute determinations of 
branching fractions at the LHC are possible using mild theoretical
assumptions), and measurements of the SM parameters. The resulting
precision for the Lagrangian parameters is compared in
\refta{tab:fittino} with a fit where only the mass measurements of
supersymmetric particles and the ratios of Higgs branching ratios at the
LHC have been used as input. \refta{tab:fittino} shows that most of the
Lagrangian parameters can hardly be constrained in a fit based on LHC
data only. The resulting parameter uncertainties often exceed the precision of
the combined fit with LHC and ILC information by orders of magnitude.
For only three of the parameters in \refta{tab:fittino}, namely
$M_{\tilde q_L}$, $M_{\tilde q_R}$ and $M_3$, the uncertainty in a fit
based on LHC input alone is in the same order of magnitude as for the
combined fit of LHC and ILC data.

\begin{table}[thb]
\renewcommand\arraystretch{1.2}
          \begin{tabular}{|l||c|c|c|c|}
            \hline
  Parameter & ``True'' value & ILC Fit value & Uncertainty & Uncertainty \\
      & & & (ILC+LHC) & (LHC only) \\
            \hline
            $\tan\beta$          &  10.00               &  10.00               & 0.11                   &    6.7 \\
            $\mu$                &  400.4  GeV  &  400.4   GeV  &
1.2 GeV     &  811.   GeV \\
            $X_{\tau}$           &  -4449.  GeV  & -4449.    GeV  &
20.GeV   & 6368.    GeV \\
            $M_{\tilde{e}_R}$    &  115.60  GeV &  115.60 GeV   &
0.27 GeV  &   39.    GeV \\
            $M_{\tilde{\tau}_R}$ &  109.89  GeV &  109.89   GeV &
0.41  GeV & 1056.       GeV \\
            $M_{\tilde{e}_L}$    &  181.30  GeV &  181.30   GeV &
0.10  GeV &   12.9    GeV \\
            $M_{\tilde{\tau}_L}$ &  179.54  GeV &  179.54   GeV &
0.14  GeV & 1369.       GeV \\
            $X_{\text{t}}$     &  -565.7  GeV & -565.7    GeV & 3.1
GeV    &  548.     GeV \\
            $X_{\text{b}}$  &  -4935. GeV & -4935.   GeV & 1284. GeV
& 6703.      GeV \\
            $M_{\tilde{u}_R}$    &  503.   GeV &   503.   GeV & 24.
GeV     &   25.   GeV \\
            $M_{\tilde{b}_R}$    &  497.   GeV &  497.    GeV & 8.
GeV      & 1269.     GeV \\
            $M_{\tilde{t}_R}$    &  380.9   GeV &  380.9    GeV & 2.5
GeV   &  753.      GeV \\
            $M_{\tilde{u}_L}$    &  523.   GeV &  523.    GeV &  10.
GeV    &   19.    GeV \\
            $M_{\tilde{t}_L}$    &  467.7   GeV &  467.7    GeV & 3.1
GeV    &  424.     GeV \\
            $M_1$                &  103.27 GeV &  103.27  GeV & 0.06
GeV    &    8.0   GeV \\
            $M_2$                &  193.45 GeV &  193.45  GeV & 0.10
GeV    &  132.     GeV \\
            $M_3$                &  569.  GeV &  569.    GeV &  7.
GeV      &   10.1    GeV \\
            $m_{\text{A}_{\text{run}}}$ & 312.0 GeV & 311.9 GeV & 4.6
GeV   & 1272.     GeV \\
            $m_{\text{t}}$     &  178.00  GeV &  178.00   GeV &  0.050
GeV  &    0.27    GeV \\
            \hline
            \multicolumn{4}{c}{$\chi^2$ for unsmeared observables: $5.3\times 10^{-5}$}\\
            \hline
          \end{tabular}
\caption{Results for Lagrangian parameters obtained in a global fit
within a variant of the SPS~1a benchmark scenario~\cite{sps}. The values in the
fourth column (``ILC $+$ LHC'') are the results of a combined fit using 
LHC and ILC data,
while the results in the fifth column (``LHC only'') 
have been obtained using LHC input only. The second column shows the
``true'' fit values, while in the third column the best fit values of
the combined fit are given (from \citere{fittino}).}
\label{tab:fittino}
\end{table}

Most of the studies of the LHC / ILC interplay in the reconstruction of
SUSY particle masses carried out so far have been done for one
particular MSSM benchmark scenario, the SPS~1a benchmark
point~\cite{sps}, since only for this benchmark point detailed
experimental simulations are available both for the LHC and ILC. 
The SPS~1a
benchmark point is a favourable scenario both for the LHC and ILC. The
interplay between the LHC and ILC could be qualitatively rather different in 
different regions of the MSSM parameter space. It seems plausible that
the synergy of the LHC and ILC will be even more important
in parameter regions which are more challenging for both colliders. 
In order to allow a quantitative assessment of the LHC / ILC
interplay also for other parameter regions, more experimental
simulations for the LHC and ILC are required.

\section{CONCLUSIONS}

The LHC and the ILC
will explore physics at the TeV scale, opening 
a new territory where ground-breaking discoveries are expected.
The physics programme of both the LHC and the ILC in exploring this
territory will be very rich. The different characteristics of the two
machines give rise to different virtues and capabilities.
The high collision energy of the LHC leads to a
large mass reach for the discovery of heavy new particles. The clean
experimental environment of the ILC allows detailed studies of directly
accessible new particles and gives rise to a high
sensitivity to indirect effects of new physics.

Thus, physics at the LHC and ILC will be complementary in many respects. 
While qualitatively this is obvious, a more quantitative investigation 
of the  interplay between the LHC and ILC requires detailed
information about the quantities that can be measured at the two
colliders and the prospective experimental
accuracies. Based on this input, case studies employing realistic
estimates for the achievable accuracy of both the experimental measurements 
and the theory predictions are necessary for different physics
scenarios in order to assess the synergy from the interplay of 
LHC and ILC.

The LHC / ILC Study Group has formed in order to tackle this task. A
first working group report, summarising the initial results obtained by 
the Study Group, has recently been completed. In this article a brief
overview about results on LHC / ILC interplay has been given,
emphasising in particular new results that have been obtained after the
completion of the first report.

In order to assess the synergy of concurrent running of the LHC
and the ILC, a wide variety of possible new physics scenarios has been
investigated, including different manifestations of the
physics of weak and strong electroweak symmetry breaking,
supersymmetric models, new gauge theories, models with extra
space-time dimensions and possible implications of gravity at the TeV
scale. These studies (of what one might term ``known unknowns'') have
revealed a number of examples where direct feedback from the ILC to LHC
analyses enables more information to be extracted from the latter. 
The interplay between the LHC and ILC is a very rich field, of which only very
little has been explored so far. The ongoing effort of the LHC and ILC
physics groups in performing thorough experimental simulations for
different scenarios will enable further quantitative assessments of the 
synergy between the LHC and the ILC. 

Besides the above-mentioned studies of ``known unknowns'', a further
part of the argument for concurrent running is based on the
``unknown unknowns'', i.e.\ the ability to interpret genuinely new
phenomena that may be observed at the LHC or ILC and that go beyond any
of the standard new physics scenarios listed above. In this case, an
unexpected observation at the ILC will be interpreted as evidence of a new
underlying theory, whose predictions can then be immediately tested at
the LHC through new dedicated searches.

In summary, experience from the past backed up by dedicated studies
using the currently most popular new physics models indicates that
concurrent running of the LHC and the ILC will significantly extend the
physics potential of both machines. The intricate interplay between
them during concurrent running will enable optimal use to be made of
the capabilities of both machines in disentangling the underlying
physics in the new TeV-scale territory that lies ahead of us. This
information will not only sharpen the goals for a subsequent phase of
running of both the LHC and ILC, but will also be crucial for determining
the future roadmap of particle physics, including of course the
subsequent generation of experimental facilities. 
A continuing effort of the LHC / ILC
Study Group is necessary in order to turn qualitative arguments into
quantitative case studies, providing in this way the results required
for a well-founded judgement of the prospective physics gain from
concurrent operation of the LHC and ILC.

\begin{acknowledgments}
The author thanks the members of the LHC / ILC Study Group
for their collaboration.
\end{acknowledgments}

\end{document}